# Optical Multicast Routing Under Light Splitter Constraints

Shadi Jawhar, Bernard Cousin
University of Rennes 1 – IRISA
Campus universitaire de Beaulieu
Rennes, F35042 France

*Abstract:* In network communications, there are many methods to exchange packets to deliver data from sources to destinations. During the past few years, we have observed the emergence of new applications that use multicast transmission. Multicasting requires "multicast routing", to deliver packets to a "group" of hosts, scattered throughout the Internet. For a multicast routing algorithm to be applicable in optical networks, it must route data only to group members, optimize and maintain loop-free routes, and concentrate the routes on a subset of network links.

For an all-optical switch to play the role of a branching router, it must be equipped with a light splitter. Light splitters are expensive equipments and therefore it will be very expensive to implement splitters on all optical switches. Optical light splitters are only implemented on some optical switches. That limited availability of light splitters raises a new problem when we want to implement multicast protocols in optical network (because usual multicast protocols make the assumption that all nodes have branching capabilities)

Another issue is the knowledge of the locations of light splitters in the optical network. Nodes in the network should be able to identify the locations of light splitters scattered in the optical network so it can construct multicast trees. This means that any node must be able to identify if other nodes in the network are multicast capable optical cross connect MC-OXC or multicast incapable ordinary cross connect OXC. These problems must be resolved by implementing a multicast routing protocol that must take into consideration that not all nodes can be branching node. As a result, a new signaling process must be implemented so that light paths can be created, spanning from source to the group members.

## I. Why Multicasting Over Optical Networks

Internet transport infrastructure is moving towards a model of high-speed routers interconnected by optical core networks. As a result, a consensus has emerged in the industry on utilizing IP-based protocols for the optical control plane. On the other hand, the rapid evolution of optical technologies makes it possible to move beyond point-to-point WDM transmission systems to an all-optical backbone network that can take full advantage of the available bandwidth by eliminating the need for per-hop packet forwarding. Multicasting over optical networks implemented with a label switching control plane, can benefit from the multicast reduction of traffic, the optical network transmission speed, and the quality of service produced by label switching.

On the other hand, constructing multicast light trees can offer an additional option that is not found in point to point light switching mechanism. GMPLS makes it possible to tunnel a set of MPLS label switched paths when they have the same ingress and egress. Moreover, multicast light trees allow grooming and tunneling of a number of point to point LSPs regardless of the egress of each.

## II. How Multicasting over Optical Networks

The multicast IP protocols had been a good area of development in the last years where many efficient algorithms were developed to be routing protocols. From this, and taking into consideration the need of compatibility between IP and non IP networks, the best mechanism to deploy multicasting principles in optical domains is to adapt the same protocols on optical communications.

Since optical cross connects and ordinary routers have a lot of differences in their packet processing, some modifications must be done to deploy the same protocols. Also, the quick evolution of "all optical switches" [10] [11], along with their efficient performance in doing switching at the physical layer, demonstrates that their use will reduce the time loss in the processing of packets from the optical layer to the network layer. Multicasting over optical networks is simply a layer 1 switching and splitting mechanism that need to be developed into algorithms and standarized into protocols.

## III. Multicast Capable OXC, Structure and Description

An OXC can switch an incoming optical signal on a certain port using a certain wavelength to an output port with the same wavelength. OXCs can be also equipped with converters that allow wavelength conversion. For an OXC to be able to switch the incoming signal into more than one output port, it needs to have the ability of branching which requires a light splitting mechanism. Light splitting mechanism allows receiving a light wave signal with a certain frequency and producing two or more light waves.

An ordinary optical node is incapable of doing splitting in the optical layer. It can only convert the light wave into an electrical signal, duplicate the electrical signal and then go back to the light layer. This is a loss of performance, and a loss of all the advantages of the all optical switch. "All-optical switch" [10] [11] can benefit from the automatic fast light switching [4] without the need of processing IP packets. A physical interpretation of the low layer switching is the wavelength division multiplexing. WDM [5] is an optical technology that provides several distinct wavelength channels on the same fiber, enabling up to 25 Tbps bandwidth [5]. When WDM networks are all-optical, they provide lower end-to-end delay by avoiding queuing and processing of packets at intermediate nodes.

Light splitters are expensive equipments based on their capability of splitting the input light wave to a certain number "m" of output light waves. Each of these m signals is then independently switched to a different output port of the OXC. Due to the splitting operation and associated power losses, the optical signals resulting from the splitting of the original incoming signal must be amplified before leaving the OXC. To ensure the quality of each outgoing signal, the maximum fan-out of the power splitter must be limited to a small integer.

If the OXC is also capable of wavelength conversion, each of the "m" outgoing signals may be independently shifted to a wavelength different than the incoming wavelength "$\lambda$". Otherwise, all m outgoing signals will be on the same wavelength "$\lambda$". Note that, just like wavelength converter devices, incorporating power splitters within an OXC is expected to increase the network cost because of the large amount of power amplification and the difficulty of fabrication. Light splitters enable the optical cross connect to split an incoming light signal from any input port to two or more output ports based on the type of the light splitter. As a conclusion, implementing standard IP multicast protocols to control multicast trees requires having light splitters on all the optical nodes of the network. Due to the highly cost of this costing assumption, some modifications need to be done so that these multicast protocols can work within optical domains even if some of the optical nodes are not multicast capable.

IV. RISING PROBLEMS

*A- Light Splitters Description*

Since multicasting is based on duplicating multiple copies of an input wavelength from an input port into several output ports, the first question that may rise is: "Are optical switches capable to split incoming light wave to more than one output light wave?" If the answer of this question is a "Yes", then a second question will definitely be "Will these optical switches perform the optical switching at optical layer, or it will go up to higher layers such as the network layer?"

Consider a WDM [5] [7] network with N nodes interconnected by fibers, each can carry W wavelengths. Each node is equipped with an OXC with P input and P output ports, some of them are multicast-capable MC-OXC. A P×P MC-OXC [9] consists of a set of W P×P splitter-and-delivery SaD switches [6], one for each wavelength. Wavelength switching is an additional capability that is very useful when multicasting is implemented. In addition to these switches, P de-multiplexers to extract individual input wavelengths and P multiplexers to combine the output ones are used. Figure 1 shows a 3 × 3 MC-OXC for 2 wavelengths.

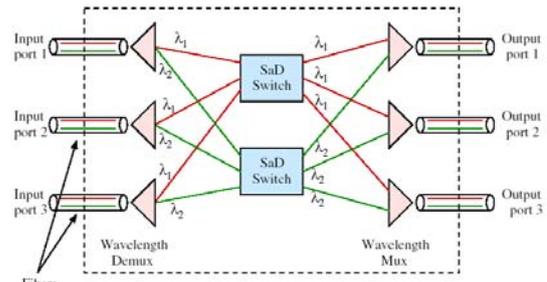

Figure 1 - MC-OXC

As shown in Figure 1, equipping an optical switch with a light splitter (to be capable of splitting input signals to more than one output signal) is an expensive issue. To reduce the cost and crosstalk on one hand, and to improve power efficiency on the other hand, a P×P SaD [9] switch which consists of P power splitters, $P^2$ optical gates, and $P^2$ photonic switching elements can be used.

We assume that the splitters are configurable to split an input into m outputs, 1<m<P (If m = 1, then there is no splitting). By configuring the corresponding m 2×1 photonic switches, each of the m resulting signals can be individually switched.

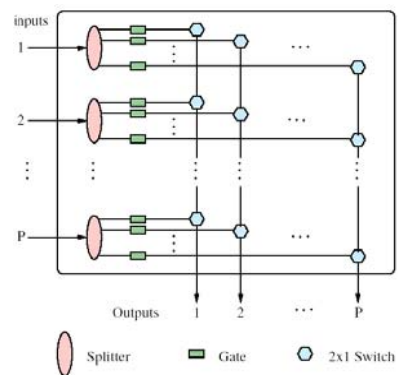

Figure 2 – SaD Switch

Even when using the above inexpensive structure, implementing splitters on all the optical switches in an optical network is a highly expensive option. Thus, a limited number of the optical switches in the optical domain will be equipped with light splitters, and therefore capable of providing the branching capability. Another important issue is the location of the light splitters and their distribution with respect to the source and destinations. Intelligence in distributing light splitters will simplify the multicast trees computation.

*B. Power Loss Problem*

The m-way splitter is an optical device which splits an input signal among "m" outputs, and therefore the power of each output is $1/m^{th}$ of the original signal. Amplification may partially compensate for power loss due to splitting, however it will raise quality of signal and noise affect problems.

## V. LIGHT SPLITTERS PROBLEM: DISCOVERY & RESOLUTION

### A. Detecting the "Unavailability of Light Splitter" Problem

The output of the multicast algorithm must be a multicast tree spanning from the source to the group members, with some nodes acting as branching nodes, and others as just forwarding nodes. The main problem occurs once a certain node is asked to be a branching node and it is not equipped with a splitter. The control entity of the optical node itself will play an important role in detecting the problem. The optical node will either discover the problem during data forwarding or during signaling prior to the data forwarding phase.

During data forwarding, when this node receives data from an input interface and is asked to split it into two or more output interfaces, it will realize its incapability. Smart nodes can realize the problem during the computation of the multicast tree before starting the data transmission.

### B. Exchange of Control Messages

When a node is asked to be a branching node and it does not possess any splitter equipment, it needs to exchange control messages to resolve this issue. This first occurs when this node receives two multicast join request messages from two downstream nodes on two different interfaces. Usually, a join request message is sent toward a source by a new group member when this group member wants to join the group.

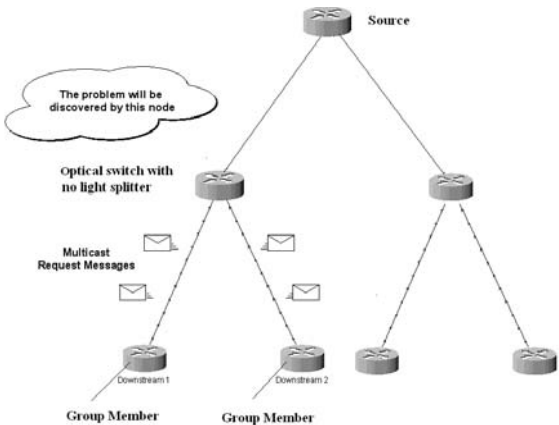

Figure 3- Exchange of control messages

As the optical node realizes its incapability of splitting, it exchange control messages to resolve this issue. Two different mechanisms can be used to update the tree. The first is based on the assumption that all optical nodes have a database that includes the locations of all the optical switches that are equipped with light splitters. The advantage of this mechanism is that the node which discovers the problem will exchange control messages in an efficient way. The second method is not based on a previous knowledge of the location of light splitters. The node which discovers the issue will have to exchange messages to find the closest splitting capable node and modify the tree. Many parameters play a role to choose the mechanism; the size of the network, the number and distribution of splitters, and the number of wavelengths. When the number of splitters and wavelengths are sufficient with respect to the size of the network, then the second method is more recommended.

### C. Control Messages with Previous Knowledge of Splitters Distribution

As shown in Figure 4, when the "optical node with no light splitter capability" receives the first join request message from the first downstream node, it will act normally, and forward this message upstream till it reaches the source, or any upstream light splitter in the multicast tree. When this node receives the second join request message, from the second downstream node, it automatically realizes the problem.

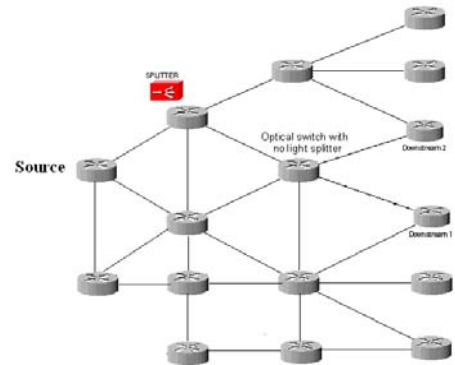

Figure 4- MC-OXC and non MC-OXC distribution

This node will use its local database to find the closest node equipped with a light splitter and simply sends a new Type 1 control message (encapsulating the second join request message) to this branching node. Two cases will appear here, and in both, the branching node with the light splitter can be an efficient participant of the multicast tree.

In the first case, the branching node is already a node of the multicast tree being constructed. In this case, the branching node needs just to save its status; it must replicate the appropriate wavelength to the newly attached downstream.

In the second case, the branching node is not a member of the multicast tree being constructed. It considers this case as the first request and forwards it upstream to the source. It identifies the source from the control message of Type 1 which includes a field with the source address. (Also, it must save its state so that it will take no other actions when a new member will request to join the group, assuring a loop free algorithm.)

The path that the second join request message had followed may not be the optimal. This occurs when the branching node is unfortunately not located between the node which has discovered the problem, and the source.

In all cases, this issue will simply be resolved by asking the branching node to send a new control message of Type 2, to the downstream node. This message asks the downstream node to send a new join message directly to the branching node and not to the source.

## D. Control Messages without any Knowledge of Splitter locations

This mechanism requires no knowledge of the locations of light splitters. When a node, with no available splitting capability, receives a second join request message, it realizes the problem. This node being a part of the current tree has now to play the role of a branching node, but it is not capable.

To resolve this problem, this node needs to find a close light splitter and forward a request to it. If it finds a light splitter which is already a node of the multicast tree, this will simplify the actions that need to be taken later. This node sends a new control message of Type 3 toward the source. This message is processed by the intermediate nodes on the path to the source. The first branching node in this sequence of nodes processes the Type 3 control message and configures its light splitter as to split the appropriate wavelength toward downstream member.

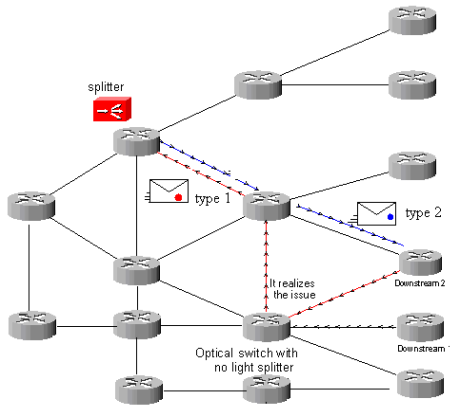

Figure 5- Discovering the problem

The simplicity of the solution is balanced with a complexity in the distribution of the locations of the optical light splitters between all the optical nodes in the network domain. This can be reduced by requiring a lesser knowledge exchange as each node requires only information about the closest light splitters.

## *Algorithm:*

| Event: A node receives a multicast join request message. |
|---|
| Reaction
(If the optical node possesses a light splitter it reserves the convenient wavelength and discards the new request.)
If the optical node possesses no light splitter, it looks into its database to locate the closest splitting node. It sends a Type 1 message (containing the request) to this splitting node. |

| Event: A multicast capable node receives a Type 1 message |
|---|
| Reaction
If this node is already a node of the tree, a message of Type 2 is sent to the group member to recalculate a best path.
If this node is not a node of the tree, it considers this as a first request and sends a join request message upstream. |

| Event: A joining member receives a Type 2 message. |
|---|
| Reaction
The group member identifies the multicast group and the branching node from the content of the Type 2 message, and sends a new join request message to the branching node. |

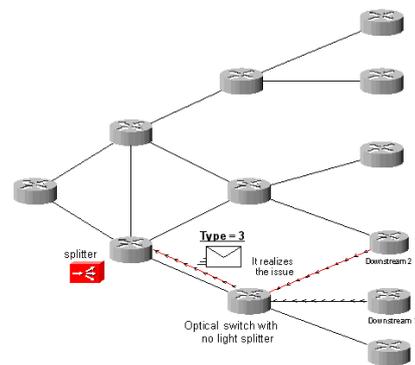

Figure 6 – locating the first splitter on the path to the source

The above scenario is ideal if the sequences of nodes from the node up to the source, contains a multicast capable node, and if the distance between them is not large. This will enhance the efficiency of the multicast tree. A TTL (time to live) field is located in the control message of Type 3 to indicate the maximum number of hops that this message is allowed to pass through. In fact, the probability of success in this operation is related to the density of splitters into the optical network.

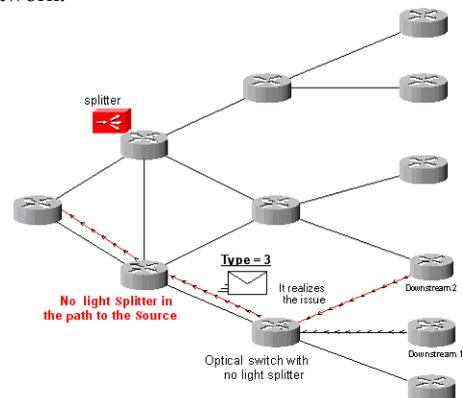

Figure 7 – locating the nearest light splitter

If the optical node that discovered the problem is unable to locate a close light splitter in the path to the source, then it tries to find a close splitter on all the interfaces, not only those spanning to the source. It does so by sending a control message of Type 4 in all directions until it receives a first acknowledgement; it updates its state so that it takes no action when receiving another acknowledgment. This assures loop free algorithm.

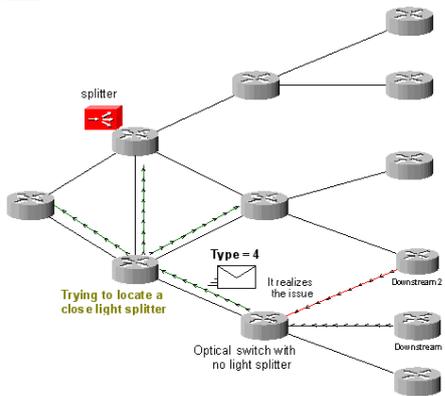

Figure 8 – locating the nearest light splitter

After it receives the first acknowledgement, it operates exactly as in the case when node depends on its database to identify the closest light splitter.

| Event: A node receives a second join request message. |
|---|
| Reaction |
| If it possesses a splitter, it reserves the convenient wavelengths and discards the request. |
| If it does not possess any light splitter, it sends a control message of Type 3 with a limited TTL toward the source. |

| Event: A multicast capable node receives a Type 3 message. |
|---|
| Reaction |
| It discards the message and configures the node as to branch the wavelength to the newly joining downstream member which is defined in the Type 3 message. |

| Event: The node does not receive any reply from any multicast capable node |
|---|
| Reaction |
| It sends a new message of Type 4 in all directions until it receives acknowledgement. (however the number of Type 4 message transmissions is bounded) |

| Event: A multicast capable node receives a reply for message of Type 3 or Type 4 |
|---|
| Reaction |
| The node locates the closest splitting node, it sends a new control message with Type 1 (containing a join request) to this splitting node. |

| Event: A multicast capable optical node receives a control message with Type 1. |
|---|
| Reaction |
| If this multicast capable node belongs to the multicast tree, it sends a message of Type 2 to the group member (to recalculate a best path). |
| If this multicast node is not a node of the multicast tree, it considers this as a first join request, and sends a join request message upstream. |

| Event: A joining group member receives a control message with Type 2. |
|---|
| Reaction |
| The group member identifies the multicast group and the branching node from the Type 2 message, and sends a join request message to the branching node. |

### E. Member Leaving the Multicast Session

When a group member requests to leave the group, it sends a "pruning request message" upstream toward the source. This request will passes through the MC-OXC, or reaches the source without passing by any branching node in the intermediate path. In one hand, actions are taken by the MC-OXC itself. It does not forward the pruning message upstream, and needs only to change its configuration in a way to stop sending the wavelength to the member.

On the other hand, if no branching node receives this pruning request message, then the convenient action will be taken by the source itself. The source forwards the pruning request message on the multicast tree so that it will be received by all branching nodes in the multicast tree. In this way, one of these branching nodes is splitting the wavelength to deliver it to the pruning member. This branching node needs to adjust its splitting configuration to stop sending selected wavelength to this branch. Then it sends an acknowledgment to the source indicating that the member is pruned. If the source does not receive any indication that the member is pruned, then it stops sending data on this branch. The result of this is to adjust the light splitting configuration on the multicast branching node or on the source itself.

In all cases, the reconfiguration of the splitting parameters will not affect the multicast tree, as the source will have the ability to continue transmitting data on the appropriate wavelength to the group members even during the pruning process.

### VI. EVALUATION OF PERFORMANCE

The performance of this proposition depends basically whether network nodes knows about the number and locations of the light splitters in the optical network, or not. From this point forward, the simulation is done separately for each of these two cases. Moreover, we separated the simulation whether there exists a multicast capable optical node, or there exists no light splitter in the path spanning from the members to the source. Different network topologies are simulated in order to simulate different sizes and structures to have general results and not only customized for special cases. Below are the table and graph of the simulation results.

| Number of Nodes: | 12 | 36 | 50 | 100 |
|---|---|---|---|---|
| Cost for building the multicast tree in case all nodes are multicast capable | 3 | 7 | 11 | 15 |
| Cost for building the multicast tree in case of knowledge of light splitters & light splitter existence in the path from members to source | 4 | 10 | 14 | 18 |
| Cost for building the multicast tree in case of no knowledge of light splitters & light splitter existence in the path from members to source | 4 | 10 | 14 | 18 |
| Cost for building the multicast tree in case of knowledge of light splitters & no light splitter existence in the path from members to source | 6 | 12 | 15 | 22 |
| Cost for building the multicast tree in case of no knowledge of light splitters & no light splitters existence in the path from members to source | 7 | 14 | 17 | 30 |

The simulation results shows that the cost for the computation of the multicast tree when the node knows have the knowledge of light splitters is exactly identical to the cost if no knowledge is available, this happens only when there exists a light splitter in the path from members to the source. In contrary, if no light splitter exits in the path from the members to the source, the knowledge of light splitters' locations shows some decreasing in the cost.

The criterion whether to have knowledge of the locations of the light splitters in the network or not, depends on the number of multicast capable optical nodes with respect to the total number of nodes. When the number of splitters is sufficient enough compared to the total number of nodes, there is a high probability to have a splitter in the path between the source and the group members, or at least one close to the node that discovers the problem. In this case, the construction of the multicast trees will be simplified and the cost of creating the multicast tree will not be negatively affected by the lack of knowledge of location of the light splitters.

VII. CONCLUSION

Our proposition performs well to resolve the "unavailability of light splitter" problem, i.e. when a multicast node is asked to be a branching node and it has no light splitter. In fact, the efficiency of our proposition is very high when the number of splitters is sufficient, and when their distribution is done in an intelligent way. Simple simulation of the amount of control messages exchanged is done, and result indicates efficiency in small domains when the number of optical light splitters is sufficient comparable to the total number of nodes in the domain. (More simulations are in progress to be able to specify the basic structure of the control messages and the necessary fields available). Also, some new fields need to be added in the control messages to resolve the issue of the power loss discussed at the beginning of this document. (Advanced simulations will result also in indicating the minimum number of light splitters required to be distributed in the optical domain, the optimum way of distributing these light splitters and finally a good measure of the efficiency of these control messages exchange in measure of time and speed).

VIII. REFERENCES

[1] R. Braudes and S. Zabele, "Requirements for Multicast Protocols". IETF RFC 1458, May 1993.

[2] A. Mankin, A. Romanow, S. Bradner, V. Paxson, "IETF Criteria for Evaluating Reliable Multicast Transport and Application Protocols". IETF RFC 2357, June 1998.

[3] G. Bourdon, "Extensions to Support Efficient Carrying of Multicast Traffic in Layer-2 Tunneling Protocol (L2TP)". IETF RFC 4045, April 2005.

[4] C. Diot, B. Neil Levine, B. Lyles, Hassan Kassem, Doug, "Deployment Issues for the IP Multicast Service and Architecture". IEEE Network. 2000.

[5] M. R. Salvador, S. M. Heemstra de Groot, D. Dey, "An All-Optical WDM Packet-Switched Network Architecture with Support for Group Communication". International Conference on Networking, 2001.

[6] W. S. Hu. and Q. J. Zeng, "Multicasting Optical Cross Connects Employing Splitter-And-Delivery Switch". IEEE Photonics Technology Letters, vol 10 pp 970-972, July 1998.

[7] C. Qiao, M. Jeong, A. Guha, X. Zhang, J. Wei, "WDM Multicasting in IP over WDM Networks". International Conference on Network Protocols, 1999.

[8] G. N. Rouskas, H. G. Perros, "A Tutorial on Optical Networks", 2002.

[9] Y. Xin, G. N. Rouskas, "Multicast Routing Under Optical Layer Constraints". INFOCOM 2004.

[10] W.-Y. Tseng, S.-Y Kuo, "All-optical multicasting on wavelength routed WDM networks with partial replication". International Conference on Information Networking, 2001.

[11] K. D. Wu., J.Y. Wei, C. Qiao, "Multicast Routing with Power Consideration in Sparse Splitting WDM networks". International Conference on Information Networking, 2001.